\input{psfig.sty}

\documentclass[12pt,preprint]{aastex}       

\begin{document}

\title{BeppoSAX serendipitous discovery of the X-ray pulsar SAX~J1802.7--2017}

   \author{G. Augello\altaffilmark{1}, R. Iaria\altaffilmark{1}, 
      N. R. Robba\altaffilmark{1}, T. Di Salvo\altaffilmark{1,2}, 
      L. Burderi\altaffilmark{3}, G. Lavagetto\altaffilmark{1}, 
      L. Stella\altaffilmark{3}}

   \altaffiltext{1}{Dipartimento di Scienze Fisiche ed Astronomiche, 
             Universit\`a di Palermo, via Archirafi 36 - 90123 Palermo, Italy;
           email: augello@gifco.fisica.unipa.it, iaria@gifco.fisica.unipa.it,
                  robba@gifco.fisica.unipa.it}
     \altaffiltext{2}{Astronomical Institute "Anton Pannekoek," University of 
              Amsterdam and Center for High-Energy Astrophysics,
              Kruislaan 403, NL 1098 SJ Amsterdam, the Netherlands;
           email: disalvo@science.uva.nl}
     \altaffiltext{3}{Osservatorio Astronomico di Roma, via Frascati 33, 
              00040 Monteporzio Catone (Roma), Italy;
           email: burderi@mporzio.astro.it, stella@mporzio.astro.it}

\begin{abstract}
  
  We report on the serendipitous discovery of a new X-ray source, 
  SAX~J1802.7--2017, $\sim 22^{'} $ away from the 
  bright X-ray source GX 9+1, during a BeppoSAX observation of the 
  latter source on 2001 September 16--20.
  SAX~J1802.7--2017 remained undetected in the first  
  50~ks of observation; the source count rate in the following 
  $\sim 300$~ks ranged between 0.04 c/s and 0.28 c/s, corresponding
  to an averaged 0.1--10 keV flux of $3.6 \times 10^{-11}$ ergs cm$^{-2}$
  s$^{-1}$.  We performed a timing analysis and found that 
  SAX~J1802.7--2017 has a pulse period of $ 139.612 $ s, 
  a projected semimajor axis of $ a_x \sin i \sim 70 $ lt-s, an orbital
  period of $\sim 4.6$ days and a mass function $f(M)\sim
  17 \pm 5\ M_{\odot}$. The new source is thus an accreting  
  X-ray pulsar in a (possibly eclipsing) high mass X-ray binary.
  The source was not detected by previous X-ray astronomy satellites, 
  indicating that it is likely a transient system. 

\end{abstract}


\maketitle
\section{Introduction}

High mass X-ray binaries (hereafter HMXBs) are young systems, 
and the neutron stars (NSs) that are often hosted in them usually 
have a strong magnetic 
field ($B \sim 10^{12}$ Gauss). Accretion onto these NSs 
occurs via capture of stellar wind matter and Roche-lobe overflow.
HMXBs comprise two main subgroups: a) the supergiant systems, and 
b) the Be star systems. In the supergiant systems,
the companion is of spectral type earlier 
than B2 and has evolved off the main sequence. Orbital periods are
generally less than 10 days, orbits are circular, and the mass transfer 
takes place because of the strong stellar wind from the OB star and/or 
because of ``incipient" Roche Lobe overflow. 
The Be star systems are characterized by emission
lines (mainly the Balmer series) which originate in the equatorial 
circumstellar envelope of the companion star. 
The orbital periods in these systems tend to be longer than those of group 
a) and correlate well with the pulsar spin period (Corbet 1986). 
The orbits are usually moderately eccentric. Transient activity is common 
in Be HMXBs; different types of outburst have been
observed from different sources, and occasionally also from the same source. 
Giant outbursts involve high peak luminosities, occur at any orbital phase and 
show only little (if any) orbital modulation of the X-ray flux. 
Recurrent outbursts usually involve lower peak luminosities, tend to occur 
close to periastron and sometimes show a strong modulation of the X-ray 
flux with the orbital phase (see e.g.\ Stella, White, \& Rosner 1986). 
To date, about 80 HMXBs are known; $\sim 40$ of them show periodic X-ray 
pulsations with spin periods distributed over a wide range from 69 ms to 
$\sim 24$ min (see Nagase 1989 for a review). 

While studying the atoll source GX 9+1 with BeppoSAX, we have discovered a new 
X-ray pulsator in a HMXB. We report here the results of the timing analysis 
of this new X-ray source. 

\section{Observations}

The observation of the GX 9+1 field was carried out from 2001 September 
16 02:01:30.0 (UTC) to 2001 September 20 03:00:08.5 (UTC), using the 
co-aligned Narrow Field Instruments (NFIs) on board BeppoSAX. 
These are: a Low Energy Concentrator Spectrometer (LECS; energy range 
0.1--10 keV; Parmar et al.\ 1997), two Medium Energy Concentrator 
Spectrometers (MECS; energy range 1--10 keV; Boella et al.\ 1997), 
a High Pressure Gas Scintillation Proportional Counter (HPGSPC; 
energy range 7--60 keV; Manzo et al.\ 1997), and a Phoswich Detector 
System (PDS; energy range 13--200 keV; Frontera et al.\ 1997). 
The exposure times were $\sim 60$ ks, $\sim 149$ ks, $\sim 142$ ks, 
$\sim 71$ ks, for LECS, MECS, HPGSPC and PDS, respectively. 
The circular field of view (FOV) of the LECS and MECS is $37^{'}$ and 
$56^{'}$ in diameter, respectively, while those of the HPGSPC and PDS 
are hexagonal with FWHM of $78^{'}$ and $66^{'}$, respectively. 
The LECS and MECS detectors are position sensitive counters with imaging 
capability. The position reconstruction uncertainty for MECS is $0.5^{'}$ 
in the central area of $9^{'}$ radius, and $ \sim 1.5^{'}$ in the outer
region of the FOV (Boella et al.\ 1997). The HPGSPC and PDS systems 
do not have imaging capabilities, and their data are therefore difficult
to interpret and analyse for individual sources when the FOV includes
more than one source.

\begin{figure}
\vspace{0cm}
\psfig{figure=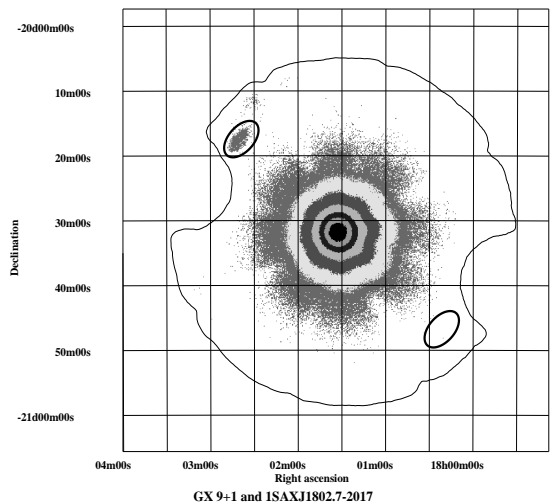,width=11.0cm,height=7.5cm}
\caption{BeppoSAX/MECS image (1-10 keV) of the GX 9+1 field. 
SAX~J1802.7--2017 is visible at the top left.
  Its angular separation from GX~9+1 is $\sim 22^{'}$. The
  extraction region of the source and background are
  also shown. The two semi-circular ``cut-outs'' are due to the removal of
  internal calibration source events. }
\label{fig1}
\end{figure}

Figure \ref{fig1} shows the BeppoSAX/MECS image (1-10 keV) centered at 
the position of the bright atoll source GX 9+1.  A fainter source is
visible at R.A.(2000.0)= 18h 02m 39.9s and Dec.(2000.0)= $-20^{\circ}$
$17^{'}$ $13.50^{''}$ (position uncertainty $2^{'}$), at an angular 
distance from GX 9+1 of $ \sim 22^{'}$.
The source was outside the FOV of the BeppoSAX/LECS. 
We have verified its presence in both the MECS2 and MECS3 images,
which probably excludes that this was a ghost image of a source outside 
the MECS FOV.
Moreover, we can be sure that the source was within 
the PDS FOV, because  the source X-ray pulsations were also detected in the 
PDS data (see below). We searched for known X-ray sources in a
circular region of $ 30^{'} $ centered at GX 9+1 in the SIMBAD data
base. We found no known sources with a position compatible with that
of the faint source; we therefore designate this serendipitous source
as SAX~J1802.7--2017.

\begin{figure}
\psfig{figure=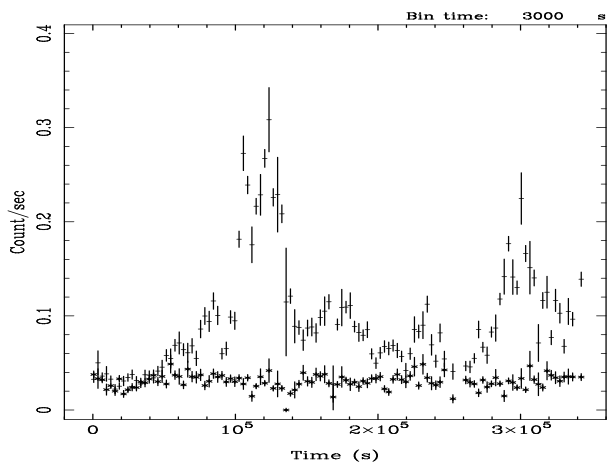,width=8.0cm,height=6.0cm}
\caption{MECS Lightcurves (1-10 keV) of SAX~J1802.7--2017 plus background 
(crosses) and of the background (stars). The bin time is 3000 s.}
\label{fig2}
\end{figure}

We extracted the MECS events of SAX~J1802.7--2017 from an elliptical
region centered at R.A.(2000.0)= 18h 02m 39.9s and Dec.(2000.0)=
$-20^{\circ} 17^{'} 13.50^{''}$ and the background events from an
elliptical region similar to the one used for the source, centered at
a symmetric position with respect to the center of the MECS FOV and
not contaminated by GX 9+1 (see Fig.  \ref{fig1}). In Figure
\ref{fig2} the lightcurves of SAX~J1802.7--2017 plus background
(crosses) and of the background (stars) are shown. The source was
not detected during the first 50 ks when the count rate within the 
extraction region ($ 0.034 \pm
0.001$ c/s) was compatible with the background count rate. 
In the following $\sim 300$~ks the source
count rate showed a large variability with an average intensity significantly
above the background level. Two flaring events took place $\sim
110$ ks and $\sim 300$ ks from the beginning of the observation.

\section{Temporal Analysis}

The arrival time of all events in the MECS and PDS
were corrected to the solar system barycenter.  We
searched for periodicities by computing a Power Spectrum Density (PSD) in
the range between $4 \times 10^{-4}$ Hz and 1 kHz from Fast Fourier
Transforms (FFTs) performed on MECS data of SAX~J1802.7--2017.

\begin{figure}
\vspace{0cm}
\psfig{figure=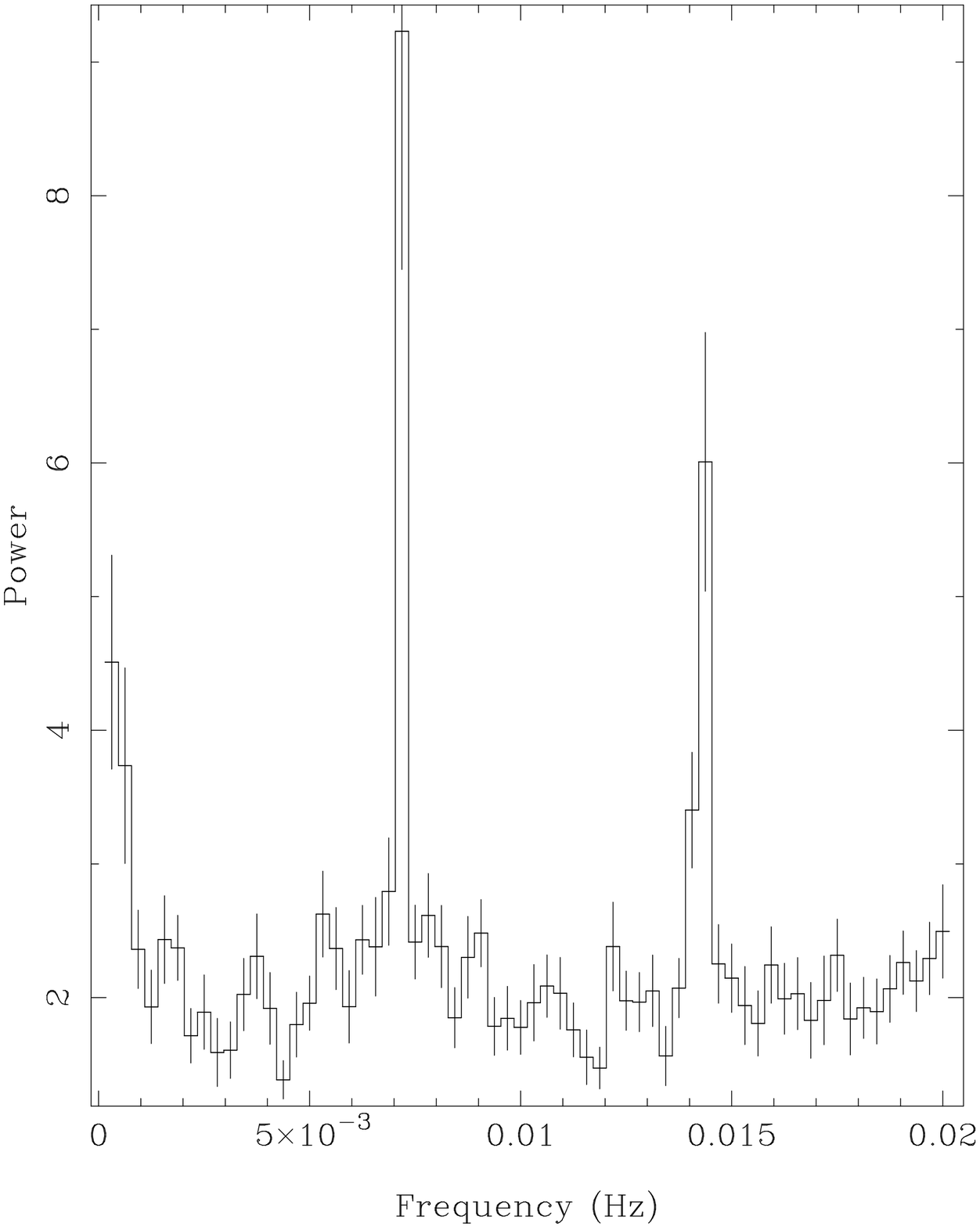,width=8.0cm,height=6.0cm}
\caption{Power spectrum density of SAX~J1802.7--2017 obtained from 
MECS data (1.0--10.5 keV 
  energy band) extracted from an elliptical region centered on the source 
  as described in the text. 
  61 PSDs, computed from time intervals of  $\sim 5 $ ks length,
  were averaged. The fundamental and second
  harmonic are clearly
  visible at 7.2 mHz and 14.4 mHz, respectively. The PSD is
  normalized according to prescription of Leahy et al.\
  (1983).}\label{fig3}
\end{figure}

A main peak at a frequency of $ \sim 7.20 $ mHz, corresponding 
to a period of $\sim 139 $ s, is evident (see Fig. \ref{fig3}).
The second harmonic is also clearly visible. We performed a folding
search for periods centered around $ \sim 139$ s finding a not well
defined period, because two $\chi^2$ peaks were present at $ \sim
139.49$ s and $ \sim 139.70$ s, respectively.

To study possible delays in the pulse arrival times, 
or, equivalently, pulse period variations we 
divided the whole data set into 35 consecutive intervals each having 
a length of about 10 ks.  We found that only 27 of these intervals had
enough statistics to carry out the analysis described below. 
A folding search was
performed in each interval for a range of trial periods 
centered around 139~s.\ The corresponding best periods were
obtained by fitting the $\chi^2$ versus trial period curve with a
Gaussian function.  The best periods varied significantly 
and with continuity between 139.44
s and 139.86 s with an average value of 139.608 s.

\begin{figure}
\vspace{0cm}
\psfig{figure=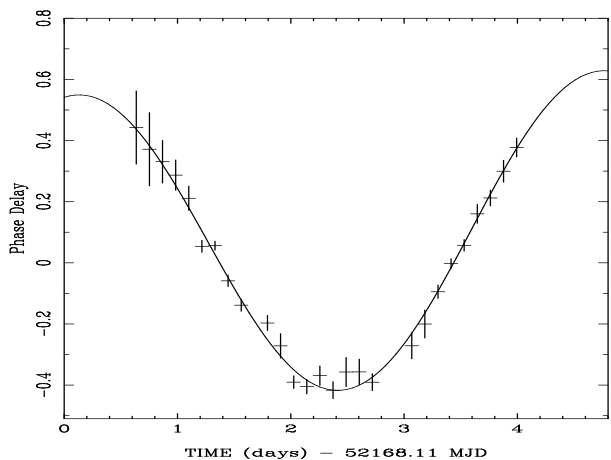,width=8.0cm,height=6.0cm}
\caption{Phase delays  of SAX~J1802.7--2017 as a function of  time. 
The solid line is the best fit function (\ref{twofor}) discussed in the
text; the orbital period is $\sim 4.6$ days. }\label{fig4}
\end{figure}

Phase delays (Fig. \ref{fig4}) were obtained by 
cross-correlating the folded lightcurves obtained for each of the 27 
intervals with that of the whole observation used as a template
(an average pulse period of 139.608 s was adopted in the folding).  
Initially we fitted the phase delays $\Delta \phi$ using
\begin{equation}
\Delta \phi = a_0 + a_1 \;t_n + a_2\;t_n^2,
\label{onefor}
\end{equation}
where $t_n$ is the arrival time of the $n$th pulse. In this formula
the linear term $ a_1=\Delta P_{pulse}/ P_{pulse}^2$ and the
quadratic term $a_2 =\dot{P}_{pulse}/2 P_{pulse}^2$ are related to 
a correction to the pulse period and the derivative of the
pulse period, respectively. We obtained a $\chi^2_{red}$ of 2.6
indicating that the modulation of the phase delays cannot only be explained
by the  presence of a derivative of the pulse period. Moreover we
find a value of $\dot{P}_{pulse}/P_{pulse}$ of $ 0.40 \pm 0.01$
yr$^{-1}$, more than an order of magnitude larger than the largest measured 
value in any known X-ray pulsar 
(i.e.\ $-10^{-2}$ yr$^{-1}$ in GX 1+4, see Pereira et al.\
1999). We checked whether the modulation of the phase delays could be
explained by the propagation delays due to the orbital motion of the 
X-ray pulsar around a companion star by fitting the phases with
\begin{equation}
\Delta \phi = a_0 + a_1 \;t_n  +B \cos{\left[ \frac{2\pi(t_n-T_{\pi/2})}{P_{orb}}\right]},
\label{twofor}
\end{equation}
The results of the fit are reported in Table \ref{Tab1}. The
$\chi^2_{red}$ was 1.3.
Since $a_x \sin i = P_{pulse} B$, we find $a_x \sin i \sim 70$ lt-s; the
corrected pulse period was $\sim 139.612$ s.

\begin{table}[h]
\caption[]{Orbital parameters of SAX~J1802.7--2017. Errors are at 1 $\sigma$
confidence level.}
\begin{center}
\label{Tab1}
\begin{tabular}{l l}
\hline \hline
Parameter & Value \\
\hline
$ a_0$ &  $ 0.04 ^{+0.12}_{-0.09} $ \\
$ a_1$ &  $ 0.02 \pm 0.03 $ days$^{-1}$ \\
$ B$ &  $ 0.50^{+0.07}_{-0.05} $  \\
$ P_{orb}$ & $ 4.6 ^{+0.4}_{-0.3} $ days \\
$ a_x \sin i$ &  $ 70^{+10}_{-7} $ lt-s \\
$T_{\pi/2}$ & $ 52168.22^{+0.10}_{-0.12}$ MJD\\ 
$ P_{pulse}$ & $ 139.612^{+0.006}_{-0.007}$ s \\ 
\hline
\end{tabular}
\end{center}
\end{table}

We tried to add a quadratic term to expression (\ref{twofor}) to take
into account a spin period derivative, but, because
of the small number of points and a relatively short observation, we
could not constrain all parameters of the fit. 
In any case, by fixing the
quadratic term in the range $-9 \times 10^{-3}$ -- $ 9 \times 10^{-3}$
days$^{-2}$ (corresponding to a $\dot{P}_{pulse}/P_{pulse}$ in the
range $-10^{-2}$ -- $ 10^{-2}$ yr$^{-1}$) we found that the orbital 
parameters do not change significantly ($\sim 2\%$ changes in the orbital
period value and $\sim 4\%$ in the $a_x \sin i$ value)
with respect to the parameters reported in Table \ref{Tab1}.

\begin{figure}
\vspace{0cm}
\psfig{figure=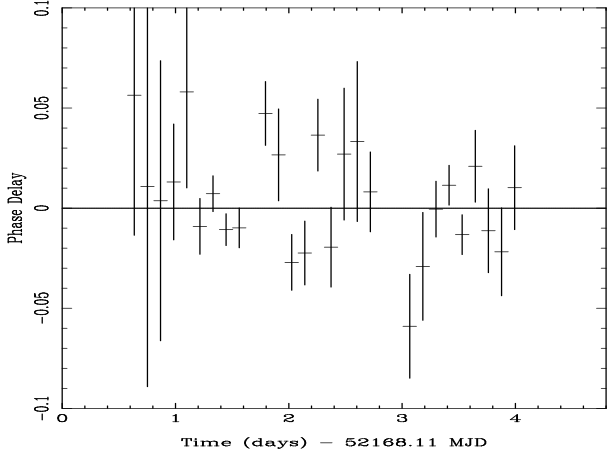,width=8.0cm,height=6.0cm}
\caption{Phase delays  after the  barycentric correction with respect 
to the center of mass of the binary system.}
\label{fig5}
\end{figure}

We corrected the arrival times of all the events, observed by the MECS
and PDS instruments, to the center of mass of the binary system using
the orbital parameters and the corrected pulse period reported in Table 
\ref{Tab1}. We then re-computed, on the corrected folded lightcurves, 
the phase delays, and found that they are compatible with zero phase delay 
(see Fig. \ref{fig5}). This suggests that our data are compatible with a 
circular orbit, although an eccentricity could be present but not 
appreciated because of the relatively low statistics of our data.
To estimate a rough upper limit to the eccentricity of the system, we 
fitted our phase delays substituting the circular orbital correction in
equation~(2) with a first-order approximation of the eccentric orbital 
correction (see e.g.\ van der Klis \& Bonnet-Bidaud 1984), thus obtaining
$e \la 0.2$ (90\% confidence level) and $\chi^2_{\rm red} \sim 1.4$.

\begin{figure}
\vspace{0cm}
\psfig{figure=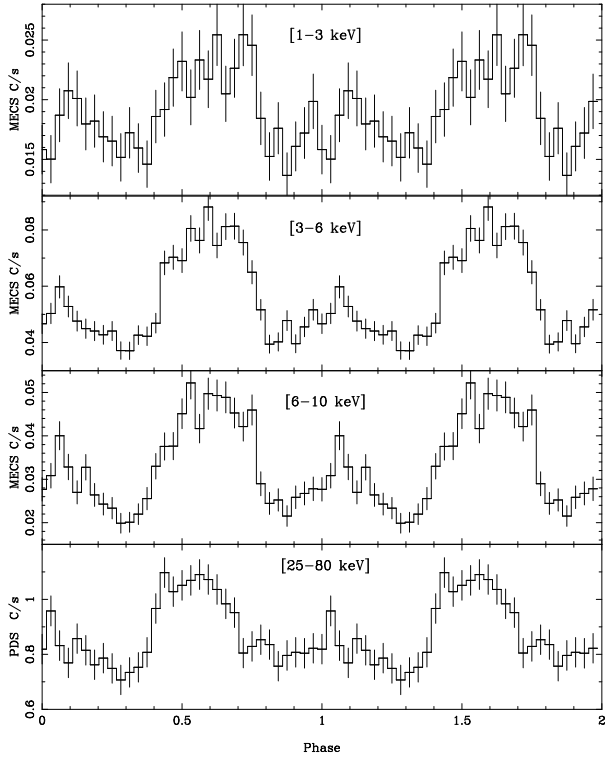,width=8.0cm,height=10.0cm}
\vspace{0.2cm}
\caption{Folded lightcurves in four different energy bands. 
From the top to  the  bottom: 1--3 keV, 3--6 keV, 6--10 keV (MECS), 
25--80 keV (PDS). The folding is obtained for a pulse period 
of  139.612 s and the  zero epoch was assumed at  the superior 
conjunction  $T_{\pi/2} = 52168.22$ MJD.}\label{fig6}
\end{figure}

Using the estimated pulse period we folded the MECS lightcurves in the
energy bands 1--3 keV, 3--6 keV and 6--10 keV, and the PDS lightcurves in
the energy bands 13--25 keV and 25--80 keV, excluding the first 50 ks of 
our observation. The main pulse of the
modulation presents a peak around phase 0.6.  A secondary peak is
visible around phase 0.1, which appears to be more prominent at  
higher energies. The pulse fraction, defined as
$(I_{max}-I_{min})/I_{max}$, with $ I_{max}$ and $ I_{min}$ the
maximum and minimum count rate, are $(43 \pm 7)$\%, $(57 \pm 6)$\%,
$(62 \pm 8)$\%, $(7 \pm 2)$\%, and $(36 \pm 3)$\% in the 1--3 keV, 3--6
keV, 6--10 keV, 13--25 keV, and 25--80 keV energy band, respectively. 
The pulse fraction in the 13--25 keV and 25--80 keV energy bands is 
strongly reduced by the presence of GX 9+1 in the PDS FOV (note however 
that the pulse fraction between 25--80 keV is somewhat higher because 
GX 9+1 is weaker at these energies). In Figure \ref{fig6} we show the 
folded lightcurves in the energy bands 1--3 keV, 3--6 keV, 6--10 keV 
and 25--80 keV.

\section{Discussion}

We have discovered a serendipitous source, SAX~J1802.7--2017, in a
BeppoSAX/NFI observation of the bright atoll source GX 9+1.
SAX~J1802.7--2017 shows coherent pulsations at a period of 139.612 s,
indicating that the compact object in this system is most likely an 
accreting magnetic NS.  Pulse arrival times show a sinusoidal modulation 
at a period of 4.6 days, which we interpret as due to the orbital motion 
of the source. 
The values of the orbital period and of the pulse period, as well as the
pulse fraction in the MECS energy range ($\sim 40\%-60\%$), are
consistent with the values usually found for HMXB systems (e.g.\ 
Oosterbroek et al.\ 1999; Bildsten et al.\ 1997). 

The mass function
of the system is $$f(M) =\frac{ M_C \sin^3 i}{ (1+q)^2} =
\left(\frac{2\pi}{P_{orb}}\right)^2 \frac{(a_x \sin i)^3}{G}\sim 17
\pm 5\; M_{\odot},$$
where $M_C$ is the companion star mass, $G$ is
the gravitational constant and $q$ is the ratio of the compact object
mass to the companion mass. The value of the mass function is not 
compatible with the system being a cataclysmic variable in which a white 
dwarf is orbiting a $\sim 1\; M_\odot$ companion.
For a NS mass of 1.4 M$_{\odot}$ we have estimated a lower limit on the
companion star mass, $ M_{C} \ga 11$ $M_{\odot}$ (90\% confidence level), 
typical of a HMXB.
The epoch of the NS superior
conjunction (which would correspond to the mid-eclipse time) derived
from the fit of the orbital parameters falls during the first 50 ks of
our observation, when the count rate of SAX~J1802.7--2017 is compatible
with the background count rate.  This is compatible
with the possible presence of an eclipse during the first 50 ks.
Assuming that the eclipse ends $ \sim 0.58 $ days after the beginning
of our observation (52168.69 MJD, see Fig. \ref{fig2}) and taking into 
account that the zero epoch is $0.11 \pm 0.10 $ days, the half duration 
of the eclipse should be $ 0.47 \pm 0.10 $ days.  
The half-angle of the eclipse subtending the portion of
circular orbit covered by the source is $ \theta = 0.64 \pm 0.14$ rad.
The eclipse duration is related to the inclination angle $i$ of the system
 by: $R_C/a = (\cos^2 i + \sin^2 i \sin^2 \theta)^{1/2}$, where
$R_C$ is the radius of a spherical primary star and $a$ is
(approximately) the separation between the two stars (see Primini et
al.\  1976).  Then a rough lower limit on the companion star radius is
$R_C \ga  14\; R_{\odot}$.

To estimate the luminosity of the
source we performed a spectral analysis in the MECS range using the
correct effective area for the position of the source in the MECS FOV
(G. Cusumano, private communication). Although the low statistics of the
data did not allow an accurate spectral analysis, we inferred a
luminosity of $\sim 5.6 \times 10^{35}$ erg/s over the 1.8--10 keV range 
by fitting the data to a power law (photon index $-0.1 \pm 0.1$ and 
$N_H = 1.7^{+0.8}_{-0.9} \times 10^{22}$ cm$^{-2}$) and assuming a source 
distance of 10 kpc.  
The lightcurve of the system shows a large variability suggesting that
the system could be accreting via a stellar wind. This is also suggested 
by its long pulse period and relatively short orbital period, typical 
of HMXBs accreting via stellar wind such as Vela X--1 and 4U 1538--52 
(see e.g.\ Corbet 1986; Nagase 1989). Indeed SAX~J1802.7--2017 shows
similarities to 4U 1538--52, which has a long spin period of $\sim 529$~s,
a short orbital period of $\sim 3.7$ days, and a mass of $19.8 \pm 3.3\; 
M_\odot$ for the companion star (Reynolds, Bell, \& Hilditch 1992).
However, while the averaged luminosity of 4U 1538--52 and Vela X--1 is 
around $5 \times 10^{36}$ ergs s$^{-1}$, the averaged luminosity of 
SAX~J1802.7--2017 measured during our observation is at least one order
of magnitude lower (assuming an upper limit to the distance of 10 kpc).
The source seems therefore to be quite underluminous with respect to 
massive binaries accreting via stellar wind, although more observations 
are needed to confirm this conclusion.

Note that, as the source has not been reported by other X-ray astronomy 
satellites before (in particular we did not find any detected source 
at the position of SAX J1802.7-2017 in ROSAT images, although the field 
of GX 9+1 was not observed by ASCA)
it might well be a transient HMXB. Most of the transient HMXBs
are Be-star systems (Liu et al.\ 2000). However, if
SAX~J1802.7--2017 belongs to this class it should be atypical
considering that its orbital period and pulse period do not follow
the Corbet correlation (Corbet 1986). 
Note also that the presence of
an eccentricity cannot be excluded by our data, and more observations 
are needed to confirm the orbital parameters of the source.

\acknowledgements
We thank Dr. Giancarlo Cusumano for providing us the corrected effective 
area for our source and the anonymous referee for useful suggestions. 
This work was partially supported by the Italian Space Agency (ASI), 
by the Ministero della Istruzione, della Universit\`a e della Ricerca 
(MIUR) and the Netherlands Organization for Scientific Research (NWO).

\end{document}